\PassOptionsToPackage{dvipsnames,table,xcdraw}{xcolor}
\documentclass[sigconf]{acmart}

\usepackage[toc,page]{appendix}
\usepackage{hyperref}
\usepackage{url}
\usepackage{graphicx}
\usepackage{booktabs}
\usepackage{graphics}
\usepackage{tabularx}
\usepackage{xspace}
\usepackage{graphicx}
\usepackage{booktabs}
\usepackage{mdwlist}
\usepackage{multirow}
\usepackage{amsmath}
\usepackage{bbm}
\usepackage{mathtools}
\usepackage{enumitem,kantlipsum} % For formal tables
\usepackage{calc}
\usepackage[neverdecrease]{paralist}
\usepackage[ruled,linesnumbered]{algorithm2e}
\usepackage{booktabs}
\usepackage{diagbox}
\usepackage{colortbl}
\usepackage{subcaption}
\usepackage{makecell}
\usepackage[figureposition=bottom,tableposition=top]{caption}
\usepackage{amssymb}% http://ctan.org/pkg/amssymb
\usepackage{pifont}% http://ctan.org/pkg/pifont

\makeatletter
\def\BState{\State\hskip-\ALG@thistlm}
\makeatother

\newcommand{\urlwofont}[1]{ \urlstyle{same}\url{#1} }

\newcommand{\eat}[1]{}

\newcommand{\hone}[1]{{\color{red} \textbf{#1}}}
\newcommand{\htwo}[1]{{\color{blue} \textbf{#1}}}

\newcommand{\vzero}{\emph{A0}\xspace}
\newcommand{\vone}{\emph{A1}\xspace}
\newcommand{\vtwo}{\emph{A2}\xspace}
\newcommand{\vthree}{\emph{A3}\xspace}

\newcommand{\nrc}{\emph{NRC}\xspace} %% #non-relevant crossed
\newcommand{\sk}{\emph{S@k}\xspace} %% success@k

\begin{document}

\title{One word at a time: adversarial attacks on retrieval models }

 \author{Nisarg Raval}
 \affiliation{\institution{Linkedin, New York}}
 \email{nraval@linkedin.com}

 \author{Manisha Verma}
 \affiliation{\institution{VerizonMedia, Yahoo! Research, New York}}
 \email{manishav@verizonmedia.com}


\begin{abstract}
Adversarial examples, generated by applying small perturbations to input features, are widely used to fool classifiers and measure their robustness to noisy inputs. However, little work has been done to evaluate the robustness of ranking models through adversarial examples.
In this work, we present a systematic approach of leveraging adversarial examples to measure the robustness of popular ranking models. We explore a simple method to generate adversarial examples that forces a ranker to incorrectly rank the documents. Using this approach, we analyze the robustness of various ranking models and the quality of perturbations generated by the adversarial attacker across two datasets. Our findings suggest that with very few token changes (1-3), the attacker can yield semantically similar perturbed documents that can fool different rankers into changing a document's score, lowering its rank by several positions. 

\end{abstract}


% \keywords{datasets, neural networks, gaze detection, text tagging}

\maketitle

\section{Introduction}
\label{sec:intro}
% False negative attacks
Deep learning ranking models increasingly represent the state-of-the art in ranking documents with respect to a user query. Different neural architectures \cite{drmm,duet,knrm} are used to model interaction between query and document text to compute relevance. These models rely heavily on the tokens and their embeddings for non-linear transformations for similarity computation. 
% The advantage of using such complex rankers and word embeddings is that they overcome several shortcomings of term-based models such as BM25\cite{bm25} which rely on exact token matches. 
Thus, they capture semantic similarity between query and document tokens circumventing the exact-match limitation of previous models such as BM25~\cite{bm25}. 

The complexity of deep learning models arises from their high dimensional decision boundaries. The hyperplanes are known to be very sensitive to changes in feature values. In computer vision literature \cite{advRank2020} and text based classification tasks \cite{wang2019survey}, researchers have shown that deep learning models can change their predictions with slight variations in inputs. There is a large body of work that investigates adversarial attacks \cite{wang2019survey} on deep networks to quantify their robustness against noise. There is, however, lack of such experiments and evaluation in information retrieval literature. Mitra \emph{et.al.}~\cite{mitra2016dualAdv} demonstrated how word replacement in documents can lead to difference in word representations which leaves room for more investigation of the robustness of deep ranking models to adversarial attacks. 

Usually, adversarial attackers perturb images to reverse a model's decision. Noise addition to images is performed at pixel level such that a human eye cannot distinguish true image from noisy image. In text classification tasks, characters or words \cite{textFool2017,papernot2016crafting} are modified to change the classifier output. 
The objective of the adversary is to reverse model's decision with minimum amount of noise injection. In information retrieval, however, it remains an open problem as to how does one design an adversary for a ranking model. We posit that the objective of an adversary in information retrieval would be to change the position of a document with respect to the query. We follow a similar approach used in text classification tasks \cite{textFool2017,papernot2016crafting} and perturb tokens in documents, replacing them with semantically similar tokens such that the rank of the document changes.

In this work, we demonstrate by means of false negative adversarial attacks that three state-of-the-art retrieval models can change a document's position with slight changes in its text. We evaluate the robustness of these ranking models on publicly available datasets by injecting varied length of noisy text in documents. We propose a black-box adversarial attack model that takes a (query, document) pair as input and generates a noisy document such that its pushed lower in the ranked list. Our system does not need access to the ranker's internal architecture, only its output given the (query, document) tuple as input. 

Our findings suggest that ranking models can be sensitive to even single token perturbations. Table \ref{tab:examples} shows some examples generated by our model of one word perturbations along with changes in document position when scored using state-of-the-art ranking model. 
% We evaluate the robustness of three ranking models on two datasets in this work. 
Overall, we found that simple attackers could perturb very few words and still generate semantically similar text that can fool a model into ranking a relevant document lower than non-relevant documents. 

%\todo{Add numbers and take-away points once we have the results}.

\begin{table}[]
    \centering
    \tiny
    \begin{tabular}{|p{1.5cm}|p{4.2cm}|l|c|}
        \hline
        \textbf{Query} &  \textbf{Relevant Document} & \textbf{Replaced by} & \textbf{$\downarrow$Rank} \\ \hline
        what can be powered by wind 
            & Wind power is the conversion of wind energy into a useful form of energy, such as using wind turbines to make electrical power , windmills for mechanical power, wind pumps for \hone{water} pumping or drainage , or sails to propel ships.
            & \htwo{wind} & 12\\ \hline
        what causes heart disease 
            & The causes of cardiovascular disease are \hone{diverse} but atherosclerosis and/or hypertension are the most common.
            & \htwo{disorder} & 8 \\ \hline
        how many books are included in the protestant Bible? 
            & Christian Bibles range from the sixty-six books of the Protestant \hone{canon} to the eighty-one books of the Ethiopian Orthodox Tewahedo Church canon.
            & \htwo{christian} & 5  \\ \hline
        how many numbers on a credit card 
            & An ISO/IEC 7812 card number is typically 16 digits in \hone{length}, and consists of:
            & \htwo{identification} & 4 \\ \hline
        how many amendments in us  
            & The Constitution has been amended seventeen additional times (for a total of \hone{twenty-seven} amendments).
            & \htwo{allowing} & 4 \\ \hline
        what is a monarch to a monarchy 
            & A monarchy is a form of government in which sovereignty is actually or nominally \hone{embodied} in a single individual (the monarch ).
            & \htwo{throne} & 4  \\ \hline
    \end{tabular}
    \caption{Examples of our one word adversarial perturbation and change in document position when ranked by DRMM.}
    \label{tab:examples}
\end{table}

\section{Related Work}
Adversarial attack on deep neural networks has been extensively explored in vision \cite{onepixel} and text classification \cite{wang2019survey,papernot2016crafting}. Existing work has proposed adversarial attacks with either white-box or black-box access to the model. In this work, our focus is on black-box attacks since access to ranking models is not always available. Existing work is also limited in that the focus is on changing classifier decisions and not document positions in ranked list with respect to a query. In this work we explore utility of (query,doc) pair in adversarial attacks on ranking models.

\begin{comment}

Main idea: Explain ranker through adversarial attacks
\begin{itemize}
    \item Small perturbation in the document significantly change its rank
    \item Can we say something about the ranker's robustness?
    \item Analyze the perturbed noise (words that need to be changed) to see if they can be used to explain the ranking.
    \item Can we take into account other documents? This way we can be different from attention based approaches that only focus on a single document.
    \item How about query perturbation? 
\end{itemize}
\end{comment}
\section{Our Approach}
\label{sec:approach}
Our goal is to minimally perturb a document such that the rank of the document changes. 
In particular, we target top-ranked documents and attempt to lower their rank through noise
injection. Assuming relevant documents are ranked higher, the objective is to lower the position of a document with minimal change in its text.
\subsection{Problem Statement}
Let $\Vec{q}$, $\Vec{d}$ and $\mathcal{F}$ represent a query, a document and a ranking model, respectively.
Given a query-document pair $(\Vec{q},\Vec{d})$, the ranker outputs a score $s=\mathcal{F}(\Vec{q},\Vec{d})$ indicating
the relevance of the document $\Vec{d}$ to the query $\Vec{q}$, where higher score means higher relevance.
Given a query and a list of documents, the ranker computes scores for every document w.r.t the 
given query and rank documents based on a descending order of the scores.
Given $\Vec{q}$ and $\Vec{d}$, our goal is to find a perturbed document $\Vec{d'}$ such that 
$\mathcal{F}(\Vec{q},\Vec{d'}) << \mathcal{F}(\Vec{q},\Vec{d})$ so that it ranks lower than
its original rank. At the same time we want to minimize the perturbation in the document.
%otherwise perturbed text can be very noisy and non-relevant with respect to the query.

We assume that both query and document are represented as vectors which could be either sequence of
terms or other suitable representations such as word embedding. Let $\Vec{q} = (q_1, q_2, ..., q_p)$
be a p-dimensional query vector and $\Vec{d} = (d_1, d_2, ..., d_q)$ be a q-dimensional document
vector. We construct a perturbed document $\Vec{d'}=(\Vec{d} + \Vec{n})$ by adding a q-dimensional noise vector 
$\Vec{n} = (n_1, n_2, ..., n_q)$. 
Our goal is to find a noise vector that reduces a document's score
without changing many terms in the document. 
% The noise vector should satisfy the following properties:
% 1) it should reduce the document's score, 
% 2) it should be sparse (should not change many terms in the document), and
% 3) it should not change the document significantly.
We formulate this problem as an optimization problem as follows:
\begin{equation}
\label{eq:v1}
\begin{aligned}
& \underset{\Vec{n}}{\text{minimize}}
& & \mathcal{F}(\Vec{q}, \Vec{d+n}) \\
& \text{subject to}
& & ||\Vec{n}||_0 \leq c 
\end{aligned}
\end{equation}
Here, $c$ is a sparsity parameter that controls the number of perturbed terms.
Although, the above formulation provides a sparse solution, we do not have
any control over the magnitude of the noise vector. In other words,
even though we change only few terms in a document we may end up
changing them significantly. To address this problem, we modify the 
objective function as follows:
\begin{equation}
\label{eq:v2}
\begin{aligned}
& \underset{\Vec{n}}{\text{minimize}}
& & \mathcal{F}(\Vec{q}, \Vec{d+n}) + \sum_{i=1}^{q}{\mathcal{D}}(d_i, (d_i+n_i)) \\
& \text{subject to}
& & ||\Vec{n}||_0 \leq c 
\end{aligned}
\end{equation}
Here, $\mathcal{D}$ can be any suitable distance function (e.g., cosine distance)
that computes distance between two terms. 
It ensures that the modified terms are close to the original terms.
We found this formulation too strict. 
In particular, it penalizes perturbations of query terms as well non-query
terms in a document. Hence, we modify the objective function to penalize
only query terms since we want to ensure that the perturbed document
stays relevant to the query. The modified objective function is as follows:
\begin{equation}
\label{eq:v3}
\begin{aligned}
& \underset{\Vec{n}}{\text{minimize}}
& & \mathcal{F}(\Vec{q}, \Vec{d+n}) + \sum_{i=1}^{q} \mathbbm{1}_{d_i \in \Vec{q}} [{\mathcal{D}}(d_i, (d_i+n_i))] \\
& \text{subject to}
& & ||\Vec{n}||_0 \leq c 
\end{aligned}
\end{equation}
Here, $\mathbbm{1}_{x}$ is an indicator function with is $1$ when $x$ is true and $0$ otherwise.
The first term in the objective function ensures that the ranker gives low score to the perturbed document
while the second term incurs penalty of changing query terms in the document. 
The constrains limits the number of changed terms in a document to $c$.

\subsection{Method}
A popular approach to solve the above problem relies on computing gradient of the model's output (score) with respect to the input and use this gradient to find a noise vector that reduces model's score on noisy input~\cite{fgsm}.
Such gradient based methods do not work well with textual data due to non-differentiable components 
such as an embedding layer. 
A common workaround is to find perturbation in the embedding space and project the perturbation vector in the input space
through techniques like nearest neighbour search. In general, gradient based methods are restricted to differentiable models and require information about the model's architecture.

To address the shortcomings of gradient based methods, we propose a different approach that uses Differential Evolution (DE)~\cite{de}, a stochastic evolutionary algorithm. It can be applied to a variety of optimization problems including non-differentiable and multimodal objective functions. DE is a population-based optimization method which works as follows:
Let's say we want to optimize over $l$ parameters. Given the population size $m$, it randomly initializes $m$ candidate solutions $\Vec{X_t} = (\Vec{x}_1, \Vec{x}_2, ..., \Vec{x}_m)$, each of length $l$. From these parent solutions it generates candidate children solutions using the following mutation criteria: 
\begin{equation}
    \Vec{x}_{a,t+1} = \Vec{x}_{b,t} + F(\Vec{x}_{c,t} - \Vec{x}_{d,t})
    % \; \; \; \; (a \neq b \neq c \neq d)
\end{equation}

Here, $a, b, c$ and $d$ are randomly chosen distinct indices in the population, and $F$ is a mutation factor in $[0,2]$. Next, these children are compared against their parents using a fitness function and $m$ candidates are selected that have the highest fitness (i.e., minimizes the fitness function). This process is repeated until either it converges or reaches the maximum number of iterations.

Given a query-document pair $(\Vec{q},\Vec{d})$ and a trained ranker $\mathcal{F}$ we use DE to find a perturbation vector $\Vec{n}$ 
that changes the document's rank without significantly modifying it.
In particular, we need to find what terms to perturb in the document and the magnitude of each perturbation. Hence, we represent the solution (perturbation vector) as a sequence of tuples $(i, v)$ where $i \in [0,|\Vec{d}|]$ represents an index of the term to perturb and $v$ represents the perturbation value. The length of the solution vector is set to $c$ (the sparsity parameter). 
We use the objective function proposed earlier as a fitness function for DE and run the algorithm for a fixed number of iterations.
Based on the choice of the fitness function, we propose three variants of the attack, \vone (Equation~\ref{eq:v1}), \vtwo (Equation~\ref{eq:v2}) and \vthree (Equation~\ref{eq:v3}). 
The final solution vector is used to construct the perturbed document $d'$. A similar approach has been used to perturb images in order to fool a classifier~\cite{onepixel}.

\begin{table}[]
    \centering
    \begin{tabular}{|c|c|c|c|c|c|c|}\hline
\textbf{Dataset} & \textbf{Model} & \textbf{\sk} & \vzero & \vone & \vtwo & \vthree  \\ \hline
\multirowcell{6}{WikiQA\\} & \multirowcell{2}{DRMM\\} & 1 & 0.62 & 1.0 & 0.93 & 1.0 \\  
                                                    & & 5 & 0.01 & 0.26 & 0 & 0.15 \\  \cline{2-7}
                           & \multirowcell{2}{Duet\\} & 1 & 0.58 & 1.0 & 0.65 & 1.0 \\  
                                                    & & 5 & 0.02 & 0.27 & 0 & 0.25 \\  \cline{2-7}
                           & \multirowcell{2}{KNRM\\} & 1 & 0.56 & 1.0 & 0.41 & 1.0 \\  
                                                    & & 5 & 0 & 0.13 & 0.01 & 0.13 \\  \hline
\multirowcell{6}{MSMarco\\} & \multirowcell{2}{DRMM\\} & 1 & 0.67 & 1.0 & 0.73 & 1.0 \\  
                                                    & & 5 & 0 & 0.11 & 0.003 & 0.04 \\  \cline{2-7}
                           & \multirowcell{2}{Duet\\} & 1 & 0.5 & 1.0 & 0.57 & 1.0 \\  
                                                    & & 5 & 0 & 0.87 & 0.03 & 0.81 \\  \cline{2-7}
                           & \multirowcell{2}{KNRM\\} & 1 & 0.58 & 1.0 & 0.89 & 1.0 \\  
                                                    & & 5 & 0.01 & 0.89 &0.45 & 0.89 \\  \hline
    \end{tabular}
    \caption{Attackers' success on changing document's position.}
    \label{table:success}
\end{table}
\section{Experimental Setup}
\label{sec:experiments}
%We evaluate the proposed adversarial attacks on three ranking models.
In this section, we provide details about the datasets, model training, perturbation and evaluation metrics. \newline
\textbf{Data:} We use WikiQA~\cite{wikiqa} and MSMarco passage ranking dataset~\cite{msmarco} for training ranking models and evaluating attacks on three ranking models.
For WikiQA, we use 2K queries for training and 240 queries for evaluation.
For MSMarco, we use 20K randomly sampled queries for training and 220 queries for evaluation. 
For each test query, we randomly sample 5 positive documents and 45 negative documents for evaluation. \newline
\textbf{Model training:} We use the following ranking models: 
DRMM~\cite{drmm}, DUET~\cite{duet} and KNRM~\cite{knrm}.
We use 300 dimensional Glove embeddings~\cite{glove} to represent each token. \newline
%All models are trained using MatchZoo\footnote{https://github.com/NTMC-Community/MatchZoo} library. \newline
\textbf{Perturbation:}
In our experiments, we perturb \emph{only} relevant documents\footnote{We found that non-relevant documents can be perturbed easily by adding noisy text.} in top 5 results for each query in the test set. 
We perturb documents using all three attackers (\vone, \vtwo, and \vthree) and evaluate ranker performance on the perturbed documents. 
The number of iterations and population size are fixed to 100 and 500 respectively. \newline
\textbf{Evaluation:}
We explore several metrics for understanding the effectiveness of perturbations and their impact on ranker performance. The goal of an   attacker is to fool the ranker into lowering the position of the document in the list. The success of the attacker is measured as the percentage of \emph{relevant} documents whose position changed by $k$ when perturbed by the attacker. Let $\mathcal{R}(d)$ denote the rank of a document $d$ and $l_d$ denote its relevance. Then, the attacker's success at k (\sk) is defined as follows:
\begin{equation*}
    \sk = \frac{\sum_{d; l_d>0} \mathbbm{1}_{(\mathcal{R}(d') - \mathcal{R}(d)) >= k}}{\sum_{d; l_d>0} \mathbbm{1}}
\end{equation*}

Here, $d'$ is a perturbation of $d$ generated by the attacker. 
We also measure the number of non-relevant documents crossed (\nrc) by perturbing $d$ to $d'$ as defined below:
\begin{equation*}
    \nrc = \sum_{\hat{d}; l_{\hat{d}}=0} \mathbbm{1}_{\mathcal{R}(d) < \mathcal{R}(\hat{d}) < \mathcal{R}(d')}
\end{equation*}
% \mathcal{R}(d) < \mathcal{R}(\hat{d}) < \mathcal{R}(d')
%Apart from these, we explore several other metrics for understanding the quality of perturbations and their impact on ranker performance.
We compare the performance of the proposed attacks against a random baseline \vzero where the adversary perturbs a word at random from the text and replaces it with the most similar word (minimum cosine distance in the embedding space) from the corpus.

\begin{table}[t]
    \centering
    \begin{tabular}{|c|c |c|c |c| }\hline
\textbf{Model} & \textbf{Atk}  & 1 token & 3 tokens & 5 tokens  \\ \hline
\multirowcell{3}{DRMM\\} &\vone&  0.944$\pm$2.776	&	3.867$\pm$6.334	&	6.814$\pm$7.858\\  
&\vtwo&  0.037$\pm$0.623	&	0.304$\pm$1.752	&	0.647$\pm$3.129\\
&\vthree&0.404$\pm$1.730	&	1.280$\pm$3.388	&	2.084$\pm$4.496\\ \hline
\multirowcell{3}{Duet\\ } &\vone&8.110$\pm$4.252	&	13.000$\pm$4.292	&	14.637$\pm$4.043	\\	  
 &\vtwo& 0.429$\pm$0.896	&	0.890$\pm$1.479	&	1.670$\pm$3.266	\\
 &\vthree& 7.187$\pm$4.284	&	10.440$\pm$4.806	&	12.055$\pm$4.771	\\ \hline
\multirowcell{3}{KNRM\\  } &\vone& 11.543$\pm$6.737	&	18.560$\pm$5.659	&	21.360$\pm$s6.459	 \\ 
 &\vtwo	&6.234$\pm$8.125	&	9.463$\pm$9.408	&	11.211$\pm$10.086\\ 
 &\vthree	&11.114$\pm$6.611	&	17.783$\pm$5.914	&	20.291$\pm$6.500\\ \hline
    \end{tabular}
    \caption{\nrc for MSMarco dataset  }
    \label{table:msmarco_nrc}
\end{table}

\begin{table}[t]
    \centering
    \begin{tabular}{|c|c |c|c |c| }\hline
\textbf{Model} & \textbf{Atk}  & 1 token & 3 tokens & 5 tokens  \\ \hline
\multirowcell{3}{DRMM\\} &\vone& 2.380$\pm$3.252 &	4.009$\pm$4.499& 4.403$\pm$4.753 \\  
&\vtwo& 0.014$\pm$0.118 &0.464$\pm$1.629 & 1.098$\pm$2.943 \\
&\vthree& 1.403$\pm$2.127 & 2.286$\pm$3.160 & 2.455$\pm$3.300 \\ \hline
\multirowcell{3}{Duet\\ } &\vone& 2.175$\pm$2.454 & 3.527$\pm$3.585 &3.773$\pm$3.857	\\	  
 &\vtwo& 	0.040$\pm$0.221 &0.175$\pm$0.506 & 0.366$\pm$0.899	\\
 &\vthree& 	2.005$\pm$2.270 & 3.175$\pm$3.277 & 3.221$\pm$3.280	\\ \hline
\multirowcell{3}{KNRM\\  } &\vone& 1.632$\pm$2.064 &2.580$\pm$2.873 & 2.849$\pm$3.155	 \\ 
 &\vtwo	&0.188$\pm$0.577 & 0.853$\pm$1.647 & 1.160$\pm$2.014\\ 
 &\vthree	&1.575$\pm$1.990 & 2.556$\pm$2.868 & 2.830$\pm$3.084\\ \hline
    \end{tabular}
    \caption{\nrc for WikiQA dataset }
    \label{table:wikiqa_nrc}
\end{table}

\section{Results and Discussion}
We focus on three research questions to investigate the impact of adversarial attacks on ranking models.\newline
\textbf{RQ1:} \emph{What is the attacker's success in changing a document's position without significantly changing the document?}
To answer this question, we measure an attacker's success (\sk) 
when we restrict the number of perturbed words to \textit{one} in each document.  The results for $k=1$ and $k=5$ are given in Table~\ref{table:success}. Notice that \vzero can change the rank of almost half of the relevant documents but not beyond five positions.
Both \vone and \vthree can change the rank of all the relevant documents by just changing \emph{one term} in the document. 
In case of MSMarco, they can lower the rank of >81\% relevant documents by more than four positions except for DRMM ranker.
Overall, we found that \vone has the highest success rate among all the attackers as it has greater flexibility to change the terms. 

\begin{figure}
    \centering
    \begin{subfigure}[b]{0.15\textwidth}
        \includegraphics[width=\textwidth]{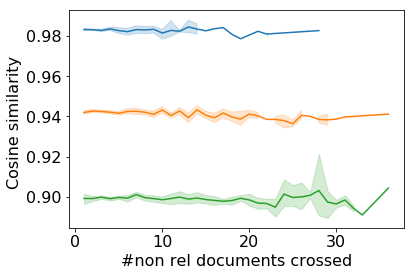}
        %{images/uniform_01_FT924-8062.png}
        \caption{DRMM (\vone)}
        \label{fig:drmm_v1}
    \end{subfigure}
    \begin{subfigure}[b]{0.15\textwidth}
        \includegraphics[width=\textwidth]{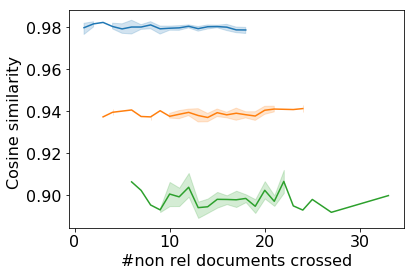}
        \caption{DUET (\vone)}
        \label{fig:duet_v1}
    \end{subfigure}
      \begin{subfigure}[b]{0.15\textwidth}
        \includegraphics[width=\textwidth]{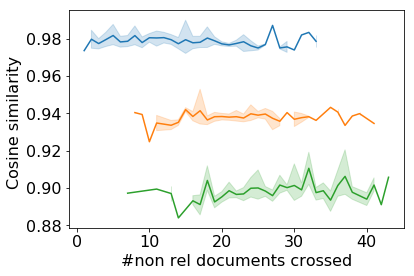}
        \caption{KNRM (\vone)}
        \label{fig:knrm_v1}
    \end{subfigure}
   \begin{subfigure}[b]{0.15\textwidth}
        \includegraphics[width=\textwidth]{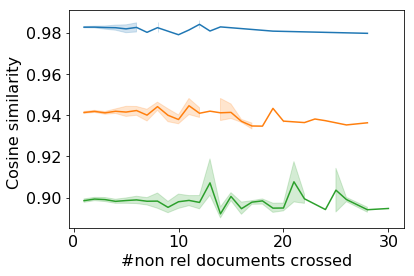}
        %{images/uniform_01_FT924-8062.png}
        \caption{DRMM (\vthree)}
        \label{fig:drmm_v3}
    \end{subfigure}
    \begin{subfigure}[b]{0.15\textwidth}
        \includegraphics[width=\textwidth]{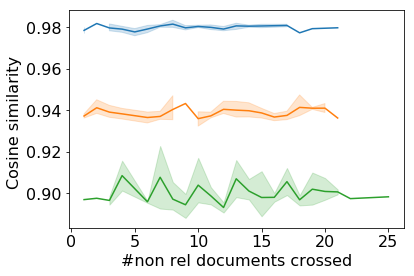}
        \caption{DUET (\vthree)}
        \label{fig:duet_v3}
    \end{subfigure}
      \begin{subfigure}[b]{0.15\textwidth}
        \includegraphics[width=\textwidth]{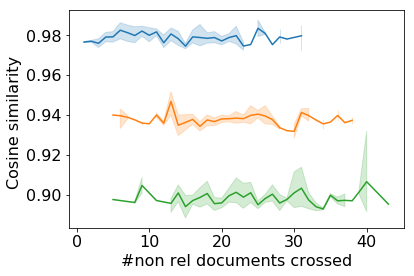}
        \caption{KNRM (\vthree)}
        \label{fig:knrm_v3}
    \end{subfigure}
\caption{\small{\nrc vs. cosine sim on MSMarco where {\color{CornflowerBlue} blue=1 word}, {\color{Peach} orange=3 words} and {\color{ForestGreen} green=5 words} perturbations respectively.}}
    \label{fig:dc}
\end{figure}

We report mean and variance of \nrc on the test sets in Table~\ref{table:msmarco_nrc} and~\ref{table:wikiqa_nrc} for MSMarco and WikiQA datasets respectively.
Increasing the number of perturbed words increases \nrc across all the attackers. On MSMarco dataset, KNRM is the most vulnerable ranker across all the attackers. On average \vone can push a relevant document beyond 11 non-relevant documents by changing only one token when evaluated against KNRM ranker. On the other hand, the performance of attackers across all the models is similar on the WikiQA dataset.
Overall, all the attackers perform better on MSMarco dataset compared to WikiQA. We argue that the larger length of passages in MSMarco provides more room for perturbations as opposed to the shorter length of answers in WikiQA. \newline
\textbf{RQ2:} \emph{What is the similarity between perturbed and original text when we restrict the number of perturbed words to $p_l=\{1,3,5\}$ in each document?} 
%We explore cosine similarity between perturbed and original text.
Adversarial perturbations may cause the meaning of document text to change. Thus, it is important to control this change such that perturbed words are semantically similar to original text. We measure the semantic similarity between embeddings of original text and perturbed text using cosine distance as done in previous work \cite{wang2019survey}. Figure \ref{fig:dc} shows similarity b/w perturbed and original text when 1, 3 or 5 tokens are changed in MSMarco passage text for three different models. 
% Token embeddings will vary across models as embedding weights were tuned, along with other parameters, for both datasets separately. 
We only focus on similarity for \vone and \vthree attackers due to space limitations. 

Overall, the cosine similarity between perturbed and original text across attackers is relatively high ($\sim$0.97), even though the document may be pushed below $\sim$20 non-relevant documents with only \emph{single} token perturbations. As expected, perturbing more tokens leads to lower similarity. Cosine similarity drops to $\sim$0.90 across models for 5 word perturbations. Both \vone and \vthree attackers can push documents to relatively lower positions with 1-3 token perturbations, however, we found that \vone tends to perturb query tokens in passages to change ranker output. We found that \vone changed query tokens in 65\% (DRMM), 11\% (Duet) and 2\% (KNRM) documents in MSMarco and  50\%, 17\% and 3\% documents in WikiQA respectively. However, \vthree achieves similar performance in terms of similarity and rank change \emph{without} changing any query tokens. \newline
%We observe similar results on WikiQA but omit them due to lack of space.
\captionsetup[figure*]{font=tiny}
\begin{table}[t]
    \centering
    \begin{tabular}{|c|c |c|c |c|c | }\hline
\multirow{2}{*}{\textbf{Model}} & \multirow{2}{*}{\textbf{Atk}}  & \multicolumn{2}{|c|}{WikiQA} &  \multicolumn{2}{|c|}{MSMarco} \\ \cline{3-6}
 &   & 1 token & 3 tokens  & 1 token & 3 tokens   \\ \hline
\multirow{3}{*}{DRMM}  &\vone &0.31$\pm$0.45	&	0.43$\pm$0.47&	0.05$\pm$0.14	&	0.11$\pm$0.17 \\  
 &\vtwo&0.005$\pm$0.06	&	0.005$\pm$0.37&	0.001$\pm$0.05	&	0.01$\pm$0.11\\
 &\vthree&0.216$\pm$0.40	&	0.30$\pm$0.44 &	0.02$\pm$0.11	&	0.06$\pm$0.15\\ \hline
\multirow{3}{*}{Duet} &\vone& 0.24$\pm$0.41	&	0.37$\pm$0.46 &	0.59$\pm$0.40	&	0.68$\pm$0.35\\
 &\vtwo& 0.003$\pm$0.07	&	-0.04$\pm$0.23 &	0.02$\pm$0.30	&	0.03$\pm$0.46 \\ 
 &\vthree& 0.226$\pm$0.40	&	0.35$\pm$0.45&	0.56$\pm$0.40	&	0.61$\pm$0.39\\ \hline
 \multirow{3}{*}{KNRM} &\vone&0.22$\pm$0.41	&	0.30$\pm$0.44&	0.53$\pm$0.37	&	0.59$\pm$0.34\\ 
 &\vtwo&0.03$\pm$0.16	&0.09$\pm$0.34&0.30$\pm$0.42&0.38$\pm$0.43 \\ 
 &\vthree&0.218$\pm$0.40	&	0.31$\pm$0.44 &	0.52$\pm$0.38	&	0.59$\pm$0.34\\ \hline
    \end{tabular}
    \caption{\% drop in $P@5$ due to perturbed text}
    \label{table:prec_drop}
\end{table}
\textbf{RQ3:} \emph{What is the ranker performance after adversarial attacks?} We evaluate model robustness against an adversarial attacker with P@5. 
% We compute perturbed $P@5$ by replacing the original document's $d_i$ ranker score $s_i$ with the new score $\hat{s_i}$ given by the ranker with the perturbed document $\hat{d_i}$ from the attacker as input.
We compute perturbed $P@5$ by replacing the original document's ($d_i$) ranker score $s_i$ with the new score $s'_i$ given by the ranker on the perturbed input $d'_i$.
\footnote{Note that ranker output for all other documents in the list  $d_{j\neq i}$ remains the same.} So, for every document the attacker perturbs, its new score replaces the old ranker score and $P@5$ is recomputed. We report the \% drop in $P@5$ for both datasets in Table \ref{table:prec_drop}. 
%\small{$^+=pval<0.10$ and $^*=pval<0.05$ with 2 tailed ttest}}
We find that \vone and \vthree attackers are able to reduce $P@5$ significantly, in some cases as much as $\sim$20\% in WikiQA and $\sim$50\% in MSMarco by changing one token in the text. Although, \vthree's drop in $P@5$ is lower than that of \vone's across all datasets since it is penalized for changing query terms. 

We also found \%drop in $P@5$ to be a function of ranker performance. Models with higher precision were harder to beat by the attacker, i.e., were more robust to token changes in document text. For example, in WikiQA original $P@5$ (mean,std) for DRMM, Duet and KNRM was 0.204$\pm$0.17, 0.207$\pm$0.18 and 0.22$\pm$0.20 respectively. It is interesting to note that all attackers are able to reduce DRMM $P@5$ by single token perturbations by highest margin. However, in MSMarco, DRMM $P@5$ is the highest amongst all other models, thus hardest to fool by all attackers. We observe very little drop in precision for \vtwo attacker across models in both datasets which indicates that very strict attackers may not be able to find suitable candidates to perturb documents. One interesting finding was that in some cases the attacker replaced document text with \emph{query tokens} to lower document score as shown in Table \ref{tab:examples}.

%\textbf{RQ3:} Finally, we explore whether relevant documents are more immune to adversarial attacks compared to non-relevant documents. The underlying hypothesis is that small changes in relevant documents should not change their position in the list drastically. 

\section{Conclusion}
\label{sec:conclusion}
Adversarial attacks on classification models both in text and vision have helped reduce the generalization error of such models. However, there is limited literature on adversarial attacks on information retrieval models. In this work, we explored effectiveness and quality of three simple methods of attacking black box deep learning models. The attackers were designed to change document text such that an information retrieval model is fooled into lowering the document position. We found that perturbed text generated by these attackers by changing few tokens is semantically similar to original text and \emph{can fool} the ranker to push a relevant document below $\sim$2-3 non-relevant documents. Our findings can be further used to train rankers with adversarial examples to reduce their generalization error.

\bibliographystyle{ACM-Reference-Format}
\bibliography{main}

\end{document}